\documentstyle[prd,aps]{revtex}
\input epsf
\begin{document}  
\draft
\twocolumn[\hsize\textwidth\columnwidth\hsize\csname
@twocolumnfalse\endcsname
\preprint{IC/97/27 
TAMU-20,
{}~hep-ph/9703465}

\title{ Neutrino - Modulino Mixing}

\author{ Karim Benakli$^{(1)}$ and 
 Alexei Yu. Smirnov$^{(2)}$}

\address{$^{(1)}$ {\it Phys. Dept.  Texas A \& M, College Station, USA. }}
\address{$^{(2)}${\it International Center for Theoretical Physics,
34100 Trieste, Italy }}

\maketitle

\begin{abstract}
We suggest an existence of  light singlet fermion, $S$, 
which interacts with observable matter only via 
Planck mass suppressed interaction: $\sim m_{3/2}/M_P$,  
where $m_{3/2}$ is the supergravity gravitino mass. 
If the mass of the singlet equals $\sim m_{3/2}^2/M_P$, then 
$\nu_e \rightarrow S$  resonance conversion solves the 
solar neutrino problem or leads to  observable effects.  
The $\nu S$-mixing  changes  
supernova neutrino fluxes and has an impact on the 
primordial nucleosynthesis. 
The singlet  $S$ can  originate as the supersymmetric partner of 
the moduli fields in supergravity 
or low energy effective theory stemming from superstrings. 
The $\nu S$-mixing may be accompanied by observable 
$R$-parity breaking effects.

\end{abstract}\vskip1pc]

{\it I. Introduction.}\hspace{0.5cm} 
Neutrinos  played  a key role in the  construction  and 
tests of the Standard model. 
It is believed that  neutrino mass, if non-zero,    
implies physics beyond the Standard model. 
We  argue that 
neutrino properties, being sensitive to 
the Planck scale suppressed 
interactions, may open  a window to hidden world.

Some time ago 
it was marked that the Planck scale 
($M_{P} = 2.4 \cdot 10^{18}$ GeV)  suppressed 
interactions can be relevant for neutrino physics 
\cite{nonren}. 
Namely,  the coupling $M_P^{-1 }L L H H $, where 
$L$ is the leptonic doublet and $H$ is the Higgs doublet 
of the Standard model,   
generates neutrino mass 
($\langle H \rangle ^2/M_P \sim 10^{-5} {\rm eV}$) 
which can lead to observable effects in solar and 
supernova neutrinos. 

There is a number 
of statements that neutrinos 
may reveal  novel  
very weak interactions with new particles. 
In last years this idea has taken rather  concrete shape. 
Observations of  the solar and atmospheric neutrinos,  
large scale structure of the Universe (the need of 
the hot component of the dark matter), LSND events {\it etc.},   
testify for non-zero neutrino mass and lepton mixing \cite{suz}. 
It is difficult to explain simultaneously  all (or even some) of  
these observations by masses and mixing 
of only three known neutrinos. In this connection    
new very light ( $m < 10$ eV)  neutral 
fermions $S$ which mix with usual neutrinos are introduced 
\cite{ster}.    
The LEP bound on the number of neutrino species,  
implies that  fermions $S$ should be  singlets of $SU(2)\times
U(1)$, {\it i.e.}  ``sterile'' neutrinos.   
Several models of the singlet fermions have been proposed recently. 
The singlet can be a component of $27$-plet in $E_6$ models \cite{ma}. 
It may have a supersymmetric origin and   
its properties may be related to the $R$-symmetry \cite{cjs1}. 
It could be a Nambu-Goldstone fermion 
in the supersymmetric theory with 
spontaneously broken global symmetry 
like lepton number or Peccei-Quinn symmetry\cite{cjs1}. 

Another suggestion is that the singlet  is a  neutrino 
from a mirror world \cite{zur}. 
The mirror neutrinos mix with usual neutrinos 
via the Planck scale suppressed interactions: 
$L^M L H^M H /M_P$, where $L^M$ and $H^M$ are the 
mirror lepton and Higgs doublets \cite{zur} 
correspondingly.

In this letter we consider new
possibilities  related to superstring theories.  

{\it II. Observation.} \hspace{0.5cm} 
A majority of extensions of the Standard 
model contain singlets of the $SU(2)\times U(1)$. 
We suggest that among these singlets there is 
at least one, $S$,   with the following 
properties: 
 
(i) $S$ has only  the Planck mass 
suppressed, $1/M_{P}$, 
interactions with the observable matter. 
In the simplest version the only light scale 
relevant  for singlets is the gravitino mass
$m_{3/2}$.  Therefore a  
dimensionless coupling constants with observable sector 
could be as 
$
\lambda = \alpha {m_{3/2}}/{M_P}~, 
$
where $\alpha = {\cal O}(1)$.  
The mixing of $S$ with  neutrinos 
involves the electroweak symmetry breaking, and  the  simplest 
appropriate effective operator is 
$\lambda \bar L S H$.   
This operator  generates  a mass term 
${m_{\nu S}}{\bar \nu }S$ with  
\begin{equation}
m_{\nu S} =  \eta \alpha 
\frac{m_{3/2}\left\langle H\right\rangle }{M_{P}} ~.
\label{mnus}
\end{equation}
Here $\eta $ 
is the renormalization effect   
and $\nu =  \nu_e, \nu_{\mu}$ or $\nu_{\tau}$.   

(ii) 
The  mass of singlet, $m_S$,  is  induced when supersymmetry is broken.   
We suggest that $m_S$  is absent at the level  
$m_{3/2}$ and appears as 
\begin{equation}
m_S = \beta \frac{m_{3/2}^2}{M_{P}}~,   
\label{ms}  
\end{equation}
where $\beta = {\cal O}(1)$. 
It turns out that  for supergravity value  
$m_{3/2} \sim 1$  TeV  
masses  (\ref{mnus}) and  (\ref{ms}) are 
in the range needed for a solution of the 
solar neutrino ($\nu_{\odot}$-) problem.  

{\it III.}  The singlet with properties 
(\ref{mnus}, \ref{ms}) has a rich {\it phenomenology}. 
Taking  $m_{3/2} = (0.1 - 3)$ TeV   and  
$\alpha$, $\beta$ in the interval 0.5 - 2 we find  
from (\ref{mnus}, \ref{ms}):  
$m_{\nu S} = (0.2 - 10) \cdot 10^{-4}~ {\rm eV}$ and 
$m_S = (4 \cdot 10^{-6} - 4 \cdot 10^{-3})$ eV.  
Manifestations of $S$  depend on  mixing angle $\theta$  with 
neutrino:
$
\tan 2\theta = 2 m_{\nu S}(m_S - m_{\nu})^{-1}, 
$    
where $m_{\nu}$ is the neutrino mass.   
From this  we find a  relation between mass 
 $m_{\nu S}$, and the  oscillation parameters, 
$\Delta m^2 \equiv m_S^2 - m_{\nu}^2$, $\sin^2 2\theta$: 
\begin{equation}
\Delta m^2 
\approx 
4 m_{\nu S}^2 
\frac{\cos 2\theta}{\sin^2 2\theta}~. 
\label{mixing2}
\end{equation}
According to  (\ref{mixing2})   
a spread of  possible values $m_{\nu S}$  
fixes region (band) of  the oscillation parameters    
$\Delta m^2, ~ \sin^2 2\theta$  
which can follow from $\nu S$ - mixing (fig. 1).  

For $m_{3/2} \sim (1 -  2)$ TeV , and  $\alpha,  
\beta \sim 1 - 2$  we get from (\ref{mnus}) and (\ref{ms})  
values $\Delta m^2$ ~$(\approx m_S^2)$ and  $\sin^2 2\theta$ 
in the range of  small mixing solution 
of the $\nu_{\odot}$-problem via the resonance
conversion  $\nu _e\rightarrow S$ in the Sun  (fig. 1) \cite{qui} . 
Notice that mixing angle relevant for 
solar neutrinos is determined by the
ratio of the electroweak scale and  the gravitino mass:  
$
\theta \sim 
{m_{\nu S}}/{m_S} \sim 
{\left\langle H\right\rangle }/{m_{3/2}}~.   
$
Forthcoming experiments, and in particular 
SNO \cite{sno},  will be able to establish whether this 
conversion takes place or not.  

\begin{figure}[t]
\centering
\epsfxsize=3.5in
\hspace*{0in}
\epsffile{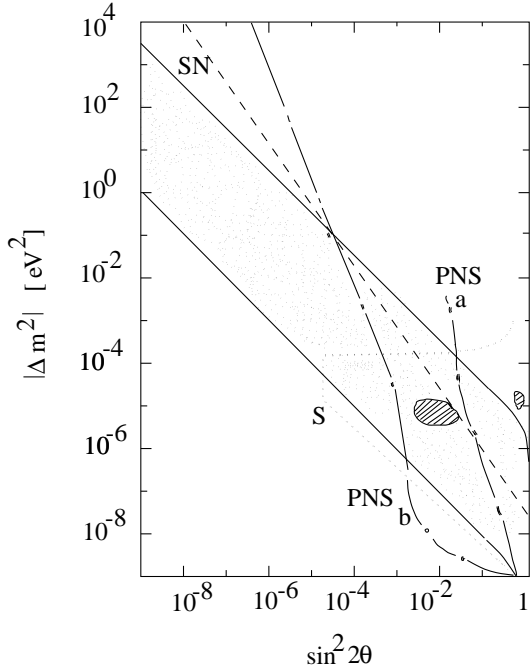}
\caption{ Oscillation parameters from the neutrino-modulino mixing 
(shadowed region). The region of solutions of the  
$\nu_{\odot}$ - problem via 
$\nu_e \rightarrow S$ resonance conversion is hatched. Also shown 
are regions 
of parameters in which neutrino-modulino mixing can be important 
for solar 
neutrinos (restricted by dotted line), for supernova neutrinos 
(dashed line shows lower edge of the region),  
and for neutrinos in the early 
universe in the epoch of primordial nucleosynthesis (PNS) 
(dashed-dotted line; (a) $m_S > m_{\nu}$, 
 (b) $m_S < m_{\nu}$ ) }
\label{fig:fig1}
\end{figure}

If the mass and mixing of $S$ are outside  the 
region of solutions of the $\nu_{\odot}$-problem, 
they still can induce an observable effect.  
A sensitivity of the $\nu_{\odot}$-data to the neutrino parameters 
is determined by the adiabaticity condition 
for the lowest detectable energy ($E \sim 0.2$ MeV):    
\begin{equation}
\Delta m^2 \cdot \sin^2 2\theta 
> 4 \cdot 10^{-10}~ {\rm eV}^2 
\label{sens}
\end{equation}
(fig. 1). This   covers the  band of 
$\nu S$-mixing (fig.1)  
for  $\Delta m^2 < 3 \cdot 10^{-4}$ eV$^2$ 
(the latter is fixed by  maximal energy of solar neutrinos and 
by central density of the Sun.)   
>From  (\ref{mixing2}) and (\ref{sens}) we  find  that 
$\nu_{\odot}$-experiments are sensitive to 
$m_{\nu_e S} > 10^{-5}$ eV.  

Let us  assume  that the  neutrino mass 
spectrum has a  hierarchy 
$m_3 \gg m_2 \gg m_1$   with 
$m_2 \sim (2 - 4)\cdot  10^{-3}$ 
eV in the range of solution of    
the $\nu_{\odot}$-problem via 
$\nu _e\rightarrow \nu_{\mu}$ conversion.  
A presence of $\nu S$-mixing will modify this  solution  
in the following way : 
  
(i). Final  neutrino flux  contains not only 
the electron and muon components but also the $S$-component. 
Moreover, the content (relative values of different fluxes) 
depends on neutrino energy.  For example, if $m_S > m_2$, 
we find \cite{BS} that flavor composition of the  final flux
can change with increase of neutrino energy as  
$(\nu_e)$ 
$\rightarrow$  
$(\nu_e, \nu_{\mu})$  
$\rightarrow$
$(\nu_{\mu}, S)$   
$\rightarrow$  
$(\nu_e, \nu_{\mu}, S)$.

Future detection of the neutral current interactions, 
and  measurements of the  ratio 
of neutral to charged current events,  
$(NC/CC)$, in different parts 
of the energy spectrum 
will allow to check  the presence of $S$-flux. 
  
(ii). A dependence of the $\nu_e$-suppression factor 
on energy (``suppression pit") is  
modified. One may expect an appearance of second pit or 
narrow dip in the non-adiabatic 
 or  adiabatic edges of the  two neutrino suppression pit \cite{BS}. 
This can be revealed in  measurements of  energy 
spectra of the boron- or  $pp$ - neutrinos. 

For $m_S < m_1 < m_{\nu_e S}$ the  
$\nu_e S$-mixing  is large,  so that  
vacuum oscillations 
$\nu_e \leftrightarrow S$ on the way from the 
Sun to the Earth become important.  
If  $\Delta m^2 \gg 10^{-10}$ eV$^2$, the  $\nu_e S$-mixing  
gives additional suppression 
of the $\nu_e$-flux by 
factor $1 - 0.5 \sin^2 2\theta_{e S}$ 
for  the energies outside 
$\nu_e - \nu_{\mu}$ suppression pit.  
For smaller values of $\Delta m^2$ 
one expects non-trivial interplay of the vacuum 
oscillations and  resonance conversion.  
If $m_S < m_{\nu S} \sim 10^{-5}$ eV,  the  
$\nu_e \leftrightarrow S$ oscillations alone 
can explain the $\nu_{\odot}$-data.  

Let us consider possible consequences of the 
$\nu S$ - mixing for the supernova neutrinos. 
Using   
density distribution $\rho \propto R^{-3}$ below the envelope of star     
($R$ is the distance from the center) 
we get from  the  adiabaticity condition  
 the sensitivity region   
\begin{equation}
\Delta m^2 \cdot \sin^3 2\theta    
> A \cdot 10^{-8}~{\rm eV}^2 . 
\label{sensSN}
\end{equation}
Here  $A \sim {\cal O}(1)$  depends on a model of  star.  
As follows from fig.1,  the $\nu S$- 
mixing can lead to appreciable transitions 
for $\Delta m^2 < 10^{-1}$ eV$^2$. This inequality 
corresponds  via the resonance condition to densities
$\rho < 10^5$ g/cm$^3$. 
Therefore $\nu S$-mixing   
does not influence  both dynamics of collapse      
($\rho > 10^8$ g/cm$^3$) 
and the supernova nucleosynthesis ($\rho > 10^6$ g/cm$^3$)  
\cite{qian}  
which occur  in the central regions of star. 
The $\nu S$-mixing can, however,  
induce  a resonance conversion in external regions of 
star thus strongly modifying  properties of neutrino fluxes  
which can be detected on  the Earth.   
If   neutrinos have  the mass hierarchy:  
$m_3 = 1 - 10$ eV, $m_1 \ll m_2 = 10^{-3} - 10^{-1}$ eV  
and  $m_S < m_1$,  the resonance 
conversion $\bar \nu_e \rightarrow \bar S$ 
will  lead to partial or complete disappearance of  the 
$\bar \nu_e$ - signal. The observation of the 
$\bar \nu_e$ signal from SN87A  allows one to put a bound 
on  $\bar \nu_e \rightarrow \bar S$ transition \cite{MS}. 
Furthermore, if the adiabaticity condition is fulfilled  in 
$\nu_{\mu} S$-resonance, the transitions  
$\nu_e \rightarrow \nu_{\tau}$ and $\nu_{\mu} \rightarrow S$ 
lead also to  disappearance of the $\nu_e$-flux.


The $\nu - S$ oscillations in the Early 
Universe generate $S$ components 
which increases the expansion rate  
and therefore influences the 
primordial   nucleosynthesis \cite{NS}. 
As  follows from fig.1, the  $\nu S$-mixing is important  
for $\Delta m^2 < 10^{-1}$ eV$^2$.  This  mixing can produce 
large leptonic asymmetry of the Universe even for larger 
values of $\Delta m^2$  which correspond to $\sin^2 2\theta > 10^{-8}$
\cite{foot}. Note that $m_S^2 - m_{\nu}^2 > 0 $ implied by 
the $\nu_{\odot}$-problem 
corresponds to weaker bound from PNS (line $a$ in fig.1) \cite{NS}. 

{\it IV. Origin of $S$. } \hspace{0.5cm} 
There are several  possibilities to get a singlet with 
desired properties 
(\ref{mnus}) and (\ref{ms}).  
The latter  imply  its  origin  in a hidden sector of the
supergravity theory. 
Superstring compactifications  lead to  
existence  of massless   singlets   
which  can be divided into two classes.

Moduli fields (dilaton, $T_i$ and  $U_i$ moduli, 
continuous Wilson lines, blowing-up modes of
orbifolds)  couple to the  observable  matter fields
only through non-renormalizable interactions 
suppressed by power of $M_{P}$. 
At string perturbative level moduli have flat potential:  
 their VEV's are unfixed and  masses are zero. 
It is believed that non-perturbative 
effects fix VEV's of moduli ( $\sim M_P$ for those having 
geometrical interpretation) and generate  masses. 
If the same non-perturbative phenomena are also responsible for SUSY
breaking, one expects that masses of modulinos are 
at most ${\cal O}(m_{3/2})$ \cite{rou}.

Non-moduli singlet fields  
can have renormalizable interactions with observable matter. 
The  compactifications typically  lead to   
several additional $U(1)'$ gauge factors (one of 
which can be anomalous)  and to a number  
of chiral supermultiplets 
singlets of standard symmetry group but  charged under $U(1)'$ 
factors.  Some of  singlets  acquire  
large ( $ \sim M_P$) VEV's  
thus breaking  $U(1)'$ factors. 
The mass matrix of the chiral fields generated as the result of this  
breaking may have small eigenvalues. 

An interesting  possibility is that 
one  of the  non-anomalous $U(1)'$ can be  broken at low scale:  
${\cal O}(m_{3/2})$ \cite{lang}, so that  
mass and mixing of $S$ (charged under this $U(1)'$) 
are protected by this symmetry.


{\it V. Singlet fermion mass.} \hspace{0.5cm} 
The supergravity mass matrix 
formula for the  fermions from (singlet)  chiral  
supermultiplets has  the form:
\begin{equation}
 M^{\alpha \beta }=
m_{3/2} N \left\langle {\cal{G}}^{\alpha \beta }  
- {\cal{G}}^{\alpha \beta \bar \gamma }{\cal{G}}_{\bar \gamma } 
+ \frac 13{\cal{G}}^\alpha
{\cal{G}}^\beta \right\rangle ~, 
\label{mass}
\end{equation}
where ${\cal{G}} \equiv  K+\ln \left| w \right|^2$,  
$K$ is  the K\"ahler potential 
and $w$ is the superpotential, 
${\cal{G}}^\alpha \equiv \partial 
{\cal{G}}/\partial \phi ^\alpha$, 
${\cal{G}}^{\bar \gamma} \equiv \partial {\cal{G}}/\partial
{\bar \phi}^{\bar \gamma}$ {\it etc.}, $N$ is  
the wave function renormalization factors  
(typically  of the order one),    
$m_{3/2}=\left\langle e^{K/2}w \right\rangle $.

The conditions for  modulino to be very light
take a  simple form if the K\"ahler function   
${\cal G} $ is written in terms of mass eigenstates.    
The singlet  $S$ should be in  the superfield  
which does not break supersymmetry, that is,   
\begin{equation}
\langle {\cal{G}}^S\rangle   = 0
\label{condi1}
\end{equation}    
(otherwise it will be eaten by the 
gravitino through the superHiggs mechanism). 
The  condition (\ref{condi1}) 
ensures  the minimum of the potential: $V^S = 0$. 
  
Using (\ref{condi1})  we can write  
a necessary condition for the mass of  
the singlet  $S$  to be  of the order $m_{3/2}^2/M_{P}$: 
\begin{equation} 
\left\langle 
{\cal{G}}^{SS} 
- {\cal{G}}^{SS {\bar \gamma} }{\cal{G}}_{\bar \gamma} 
\right\rangle 
\sim {m_{3/2} \over M_{P}}~.  
\label{condi} 
\end{equation} 
If  $S$ does  not mix with  fields which break SUSY,   
that is, 
$ 
\left\langle 
{\cal{G}}^{SS {\bar \gamma} } 
\right\rangle = 0
$
for all
$
\left\langle 
{\cal{G}}_{\bar \gamma} 
\right\rangle \neq 0, 
$ then  (\ref{condi}) is reduced to  
$ 
\left\langle 
{\cal{G}}^{SS}  
\right\rangle \approx 0,  
$ 
while usually one expects  $\left\langle {\cal{G}}^{SS} 
\right\rangle \sim {\cal O} (1)$  
\cite{rou}.

Let us consider how the conditions (\ref{condi1},\ref{condi}) could be 
implemented for some simple  K\"ahler 
potentials which are known to arise from 
string compactifications.

If $\Phi$ is one of the   moduli describing geometry of the compactified
space of orbifolds or Calabi-Yau 
  (like the  $T$ moduli) then 
$\langle \Phi \rangle \sim 1$  and  in the large volume approximation    
the  K\"ahler potential has the form:  
\begin{equation}
K = p\ln(\bar \Phi + \Phi - 
\bar z^{\bar \gamma} z^\gamma) + K_{ho}~.    
\label{kaha}
\end{equation}
Here $z^{\gamma}$  represent 
Wilson line moduli and      
matter fields.  
$K_{ho}$ stands for all  higher order corrections 
and  unknown non-perturbative  contributions.  
The fields are in units of Planck mass;   
$p$ is an integer (typically $p=-1,-2,-3$). 

We find  $Det K^{\Phi \gamma} = 0$,   
and in the case of one field $z$ 
the state with zero eigenvalue equals 
$S = \cos \alpha \Phi + \sin \alpha z$, 
where 
$\tan \alpha = - 1/  \langle z \rangle$. 
If ${\cal G} \approx K$,  then  fermion $S$ satisfying 
condition (\ref{condi}) will have zero mass.   
A finite contribution to  $m_S$ can follow 
from $K_{ho}$ or/and  non-perturbative part of the  superpotential
related to SUSY breaking.

If VEV of $z$ is small or zero  the conditions  
(\ref{condi1}) (\ref{condi}) for $S \approx \Phi$ can be satisfied by
cancellation
of contributions from the K\"ahler potential and  superpotential.  
The cancellation can be easily realized for  polynomial 
superpotentials.  In particular, 
for $w = a( S - \langle \bar S \rangle)$ the condition (\ref{condi})  
is fulfilled automatically.  
It is believed, however, that the whole theory  
is invariant under the shift
$ S \rightarrow S + i$ \cite{ds,can}. In this case the 
superpotential has a general  form  
$
w =  e^{-2\pi a S} \sum_n a_n exp (-2\pi n (S - \langle S \rangle)).  
$
If $a \approx p /2\pi \langle S + \bar S \rangle$  
the series  converges very quickly and  
we can write the superpotential explicitly as: 
\begin{equation}  
{\displaystyle 
w \approx A e^{\frac{pS}{\langle S + \bar S \rangle}} 
\left[ 1 + \frac{p}{4 \pi^2} e^{-2\pi (S - \langle S \rangle)} \right]
} ~. 
\label{suppot}
\end{equation} 
Here $A \sim m_{3/2}M_P^2$.  
For other values of $a$ the coefficients in the expansion are 
large.

The  mass of the singlet $m_S$ can be generated by second term in 
(\ref{condi}). For  the K\"ahler potential (\ref{kaha}) we get 
$
m_S = - 2p m_{3/2}
\langle z \rangle \langle {\cal G}_{\bar{z}}\rangle /
(\langle S + \bar S \rangle)^3 
$
and a correct order of magnitude is achieved  
for 
$\langle z \rangle = m_{3/2}$ 
and  
$\langle {\cal G}_{\bar{z}}\rangle  = 1$.

If the field $S$ has a small ($ \ll M_P$) or vanishing VEV,  
the K\"ahler potential can be expanded as: 
\begin{equation}
K = K^{S{\bar S}}{\bar S} S + \frac {{\bar z} z}{M_{P}^2} 
( SS + h.c.)+ K_{ho}~,
\label{kahb}
\end{equation}
and the superpotential can be {\it a priori} 
an arbitrary holomorphic function of $S$. 
The K\"ahler potential 
(\ref{kahb})  mixes $S$ with 
the  field $z$.  
If ${\cal G}^S = 0$ and 
$\langle w^{SS}\rangle = 0$ (which is easy to satisfy), 
the mass of $S$ can be written as 
$
m_S = m_{3/2} \langle 2 \bar z z - 2 z {\cal G}^z - S^2 \rangle~. 
$
Now there are different ways to get a desired
value of $m_S$: 
(i) 
$\langle z \rangle = 
\Lambda_{hid} \sim (m_{3/2} M_{P})^{1/2}$,  
$\langle {\cal{G}}_{\bar z }\rangle = 0$,    
$\langle s \rangle < \Lambda_{hid} $.    
(ii) $\langle {\cal{G}}_{\bar z }\rangle \sim 1$,   
$\langle z \rangle = m_{3/2}$. 
(iii) $\langle s \rangle =  \Lambda_{hid} $ 
(without mixing with $z$ field).

The mass $m_S$ can originate from mixing of $S$ 
with fields $\Phi$  having a Planck scale mass, provided   
the  $S \Phi$-mixing  is  the order  
$m_{3/2}$. The latter scale  can appear from the K\"ahler potential 
in the same way as  the $\mu$-term appears. 
It can be protected by additional $U(1)'$ gauge symmetry   
broken at $m_{3/2}$, if $S$ is charged under $U(1)'$,  
whereas $\phi$ is a singlet of this group. 
In this case  for the mass of $S$ we 
have  usual see-saw formula: $m_S = m_{3/2}^2/M_{P}$. 

Another possibility is when the  superfield $S$ charged under $U(1)'$
gets a VEV of the order $m_{3/2}$. This VEV  leads 
to mixing of the fermion $S$ and gaugino associated with 
$U(1)'$. If this gaugino has the Majorana mass of the 
order $M_P$, then again the see-saw mechanism results in 
 the desirable mass of $S$. 


{\it VI. Mixing. } \hspace{0.5cm}
The interaction $\bar L H_2 S$ 
with the coupling constant $ \lambda \sim m_{3/2}/M_P$  
can  be generated either through non-renormalizable 
interactions in the superpotential or from the K\"ahler potential 
in a  way similar to appearance of the $\mu $-term
for the Higgs doublets 
\cite{giud,agnt}. Let us consider the following coupling:      
\begin{equation}
K = ... +  \frac{1}{M_P} P(S) L H_2 + h.c. .. , 
\label{kahc}
\end{equation}
where $P$ is some function of moduli $S$. 
For $P(S) = S$ we find immediately the desired mixing mass: 
$
m_{\nu S} \sim m_{3/2} K^{SL} \sim   
{m_{3/2} \langle H_2 \rangle}/{M_P}~. 
$
In general, 
\begin{equation}
m_{\nu S} = 
\left( K_{S\bar S}K_{L\bar L}\right) ^{-1/2} 
\langle H_2(e^{\cal G} \tilde \mu  + m_{3/2} P^S - {\bar F}^{\bar z} 
\partial_z P) \rangle . 
\label{mu}
\end{equation}
Here $\tilde \mu _i$  is the  effective mass induced by 
non - perturbative effects or higher derivative terms \cite{agnt},  
and ${\bar F}^{\bar z}$ appears if $P = P(S, z, \bar z)$. 
${\bar F}^{\bar a}$ and $\tilde \mu _i$ are  
functions of $S$.

A generic feature of the neutrino-modulino mixing 
is the R-parity violation by
dimension three operators. Indeed, the same term of the  
K\"ahler potential (\ref{kahc}) generates the 
lepton violating coupling $m_{LH}LH_2$.  The mass 
$m_{LH}$ has a general expression (\ref{mu}).  
Performing its expansion in 
series of $S/M_{P}$ we can write: 
\begin{equation}
m_{3/2} \left[ \alpha' 
+ \beta' \frac{\langle S \rangle}{M_P} \right] 
L H_2  ~. 
\label{lepthiggs}
\end{equation} 
The terms in  (\ref{lepthiggs}) can be separately small,  
and   $ m_{LH}$ can range from zero to  
$m_{3/2}$, although for some moduli (like $T$) one expects 
$m_{LH} \sim {\cal O}(m_{3/2})$.

Explicit R-parity violation by dimension two operators 
has  interesting phenomenological consequences \cite{hall}.  
It generates  lepton number violating Yukawa couplings 
$\lambda$ and  $\lambda'$. 
Even in the case of universal 
soft symmetry breaking terms at high scales it leads 
due to renormalization group effect
to nonzero VEV for
sneutrino $\left\langle L\right\rangle \neq 0$. 
This in turn  results in mixing of neutrino and neutralinos 
and generation of masses for light neutrinos \cite{hall}.  

The present scenario is  based on 
gravity mediated SUSY breaking. Its signature is 
the neutrino transitions into singlet state 
and the $R$-parity violating effects.\\


The authors wish  to thank R. Arnowitt, C. Kounnas, 
K. Narain, J. Rizos and F. Zwirner for 
discussions. The work of K. B. was supported  by the DOE grant 
DE-FG03-95ER40917.
\\[0.2cm]

\end{document}